# Astronomy and Feng Shui in the project of the Tang, Ming and Qing royal mausoleums: a satellite imagery approach

Giulio Magli
School of Architecture, Urban Planning and Construction Engineering,
Politecnico di Milano, Italy
Giulio.Magli@polimi.it

*The royal Chinese mausoleums of the Tang, Ming and Qing Chinese dynasties are astounding ensables of monuments, conceived and built to assure to the Emperors immortality in the afterlife and perennial fame on earth. To this aim, a series of cognitive elements were embodied in the funerary landscapes selected for such monuments, including astronomy, general topography, and traditional Chinese "geomancy". Taking advantage of satellite imagery, we investigate here on this issue in a general manner. In particular, we develop and apply a rigorous approach to investigate if magnetic compass was used in the planning of such monuments.*

## 1. Introduction

Starting from the first Emperor of Qin, different solutions were adopted by the Chinese rulers for their tombs, and – consequently - spectacular royal necropolises were founded and developed. In particular, the Han (206 BC–220 AD) and later the Song (960–1279 AD) emperors opted for the construction of tombs located under huge artificial mounds. In a recent paper, we have analyzed several cognitive aspects of these impressive monuments and, in particular, we have shown that their orientation was governed by astronomical considerations (Magli 2018). In the present paper, we study the funerary landscapes of the Tang, Ming and Qing dynasties. The tombs of these emperors were not based on the construction of artificial mounds: the Tang rulers took advantage of existing mountain peaks, while the Ming and Qing tombs were conceived as architectural units above ground with a corresponding palace underground. This research aims at a better understanding of relevant cognitive aspects of all these monuments, such as orientation and placement in the landscape. Although the author has visited the vast majority of the monuments discussed, the broad approach presented here has been greatly facilitated by the use of satellite imagery tools, both for estimating orientation and for gaining a better glimpse of the local landscape surrounding each tomb. The results presented are twofold: on one side we show for the first time the role of astronomy in the planning of the Tang tombs, on the other we confirm the repeatedly supposed role of *form* Feng Shui in planning the Ming and the Eastern Qing tombs. Further, we perform a complete scientific analysis of the possible use of the magnetic compass in the planning of these monuments – and therefore of the so called *compass* Feng Shui doctrine – obtaining fully negative results, that is, we proof that magnetic compass was never used. Finally, for the special case of the Western Qing cemetery, a topographical hypothesis is proposed to explain its location.

## 2. The Feng Shui tradition(s) in brief and the methods for testing their presence

Feng Shui is the traditional Chinese pseudo-scientific doctrine of "geomancy", that is, divination in accordance with the geographical and morphological features of the terrain (Yi et al. 1994, Bruun 2008). In a nutshell, the idea is that special locations exhibiting particular features are favorable from the point of view of encapsulating and enhancing "qi", a "positive energy" allegedly flowing on the Earth, and are therefore to be preferred for city planning and especially for tombs' locations. Of course, it is worth stressing that the idea of a "terrestrial energy" has no scientific basis whatsoever; Feng Shui is however very important from the historical point of view in the contexts discussed in the present paper. Being not even a "theory" in the sense of a well established framework of rigid rules, it is not a easy task to reconstruct the history of Feng Shui. If indeed certain points, common to general traditions and beliefs of the Chinese civilization – like the identification of the four cardinal points with four symbolic animals – can be traced back up to Chinese Bronze Age or even earlier (Pankenier 2015), written records directly referring to the geomancy of the terrain do not appear up to the end of the Han dynasty. In any case, it can be asserted that Feng Shui developed along two different, although intervined, branches. For simplicity, we shall cal them with the traditional terms of *form school* and *compass school* respectively.

The form Feng Shui school developed first, and relied on observation of the morphology of mountains and rivers and of their relationship to winds. These observations led to a rather complex way to select favorable places. The main idea was that a "correct place" must first of all have the tallest visible mountain to the north and a river flowing to the south. Starting from these two main characteristics, a series of further, intricate refinements and analysis were applied to establish the supposed level of suitableness of a site. The principal refinements include the shape of the main mountain (the Black Turtle) which must be undulating (Dragon vein shape), the presence of protective hills to the west ( Blue Dragon hill) and to the east (White Tiger hill), a facing, lower hill to the south, and a meandering (curving) flow of water between the site and the south hill. The presence of dense vegetation on the mountains was considered as further, important index of auspiciousness, as well as the quality of the terrain, which had to be dry. These characteristics were further specified using a complex classification of shapes of the hills, types of river's curves, and so on. Once all (or at least most) such requirements were satisfied, the place was declared suitable for the building of the "Feng Shui cave" - the tomb. Although it may seem strange, the underlying idea was that the vital energy flowing in such an auspicious place, passing along the bones of the deceased, would have been transferred to his family assuring prosperity and health to his living descendants.

The compass Feng Shui school developed after the discovery of the magnetic properties of the lodestone, and was based on the magnetic compass, again allegedly "measuring" the above mentioned "currents" on earth. According to the authoritative account by Needham (1962) an early kind of magnetic compass was already invented in China during the Han Dynasty, around the second century BC. The instrument was a spoon-shaped piece of lodestone resting freely on the concave part and placed on a cast bronze plate, so that it could in principle orient according to the magnetic field. Of course, it was a very imprecise instrument, and indeed it was not used for navigation but only for "magic" purposes. The great breakthrough occurred much later, during the late Tang Dynasty with the invention of the freely floating (either wet or dry) magnetic needle. The Compass Feng Shui is thus soundly documented from the Tang Dynasty onwards (see again Needham 1962). In Compass Feng Shui the directions characterizing a site – flowing of rivers, orientation of hills veins, and so on – are measured and to each of them a degree of auspiciousness is assigned depending on the sector of the compass girdle indicated by the magnetic needle.

To test if form Feng Shui was used in choosing a place is – generally - not an easy task. The reason is that this doctrine is not a rigorously established one, and different authors may consider specific features of places in different way or with a different order of importance (Hong-Key 2008). Examples exist of important ancient places which are said to be especially auspicious but whose main characteristics do not match the well established Feng shui "master lines" (in particular, during the Han dynasty both towns and royal tombs were planned without any consideration of Feng Shui properties). In the present paper we shall be facilitated by satellite imagery in order to check if the main standard requirements are satisfied by a specific funerary landscape.

A much more rigorous test can be performed about compass Feng Shui, which can actually be explored scientifically. The method followed in the present paper is as follows. First of all, we must recall that the Earth's magnetic poles do not coincide with the earth's rotation poles and continuously move over the earth's surface. Therefore, although the Earth's magnetic field can be ideally visualized as a dipole positioned at Earth's center and aligned with the rotational axis of the Earth, in practice the direction of the magnetic compass bearing does not correspond to the geographical north and varies continuously. The difference in degrees between these two directions at a given place and instant of time is called magnetic declination. It may happen that the magnetic declination is close to zero, but it may also happen that it deviates macroscopically (up to ten degrees or even more).

The record of the strength and direction of Earth's magnetic field (paleomagnetism) during geological times is very important and can be done by studying the magnetic properties of rocks. Quite a different problem is, instead, the reconstruction of the magnetic field in historical times. In recent years however, quite affordable models of the variation of magnetic declination have been produced. In the present paper we have used the model CALS10k.2 of the Earth's Magnetic Field developed at German Research Centre for Geosciences at Potsdam. The model returns the expected magnetic declination at a given time and at given geographical coordinates (Constable, Korte and Panovska 2016).

As a first step, we thus calculated the values of the magnetic declination at the geographical coordinates of the tombs in question. As reference times we took the year of accession of each emperor, since it is well known that planning of the ruler's tomb initiated very early in his reign, usually in the first year. As a second step, we measured orientation of each tomb (details of the elements measured are given in each section). For the sake of internal coherency all data reported in the present paper have been obtained trough satellite imagery extracted from Google Earth (or Bing) and imported in AutoCad. In all cases the measurements have been repeated on several images of the same place taken from the historical archives of the programs. On account of the high quality of the images and of the low projection error associated with them, the intrinsic error expected from this kind of measurement is quite low (Potere 2008). In any case, sample checks of some of the data in each table have been made trough other means, mostly with precision compass readings on site by the author but also using existing surveys. The results of these tests confirm the validity of the satellite imagery approach.

Once the magnetic data and the orientation data have been obtained, we proceeded using a test of correlation. A few things are worth stressing here. If indeed the compass was used to orient buildings, then there must obviously be a *linear* correlation between magnetic declination and orientation. Therefore, it is sufficient to test correlation using the so-called Pearson correlation R. If the R-test fails, then there is no correlation at all (a thing which in a general case would be untrue, since a non-linear correlation between two sets of data would escape the test). Pearson's correlation coefficient is the covariance of the two variables under exam divided by the product of their standard deviations. It is therefore a number between +1 and −1. Value close to 1 indicates a positive, strong linear correlation, values close to -1 a negative, strong linear correlation, while values close to 0 show that the two variables are not linearly correlated. Again, due to the special nature of our data, in our case negative R-values eventually occurring would also show that no correlation exist at all. Visually, sample of data generating R-positive results typically are recognizable at simple inspection, because highly unrelated data always appear as "clouds" in the X-Y plane. Therefore, for each set a visualization of data has been also performed. Of course, the orientation measured in azimuth (between 0° and 360°) have been converted in the same format of the magnetic declination data – positive if towards the east, negative otherwise – before performing the analysis.

# 3. The Tang dynasty mausoleums

Starting with the first emperor of Qin, and in the course of the Western Han dynasty, the Chinese emperors ordered the construction of their tombs under huge burial mounds in the form of square-base "pyramids". The same tradition was to be revived by the (Northern) Song Dynasty (960-1127 AD).
Among the various dynasties in between the Han and the Song, the Tang (618–907 AD) period singles out. In particular, under the Tang, a new architectural paradigm was introduced (Zhou 2008). Sources indeed report that the first Tang emperor Taizong ordered the construction of his father's mausoleum according to the Han traditions. The tomb was thus built on the ridge of the Xi'an flatland, to the north of river Wei, and is characterized by a huge, cardinally oriented mound. However, regarding his own tomb, the Emperor explicitly stated that it had to be inspired by Baling, the tomb of Emperor Wen of Han, the unique Han ruler who built its tomb under a natural mountain. The Taizong mausoleum, Zhaoling, was therefore located under a mountain, Mount Jiuzong. The choice of this mountain was very careful: it is a peak located to the northeast of Liquan County, Shaanxi Province. The peak has a a very peculiar, unmistakable profile, and ancient sources report that the mountain was sufficiently high to be seen - as a small heap at the horizon - from the Tang capital, today's Xian. Of course today this sight would be almost impossible, but it may well be that it was possible in ancient times. Indeed the relative height is around 600 meters and the distance from Xian centre around 55 Kms, while the theoretical visibility distance in Kilometers (the square root of 13 times the height in meters) is of the order of 88 Kms. Interestingly enough, satellite imagery shows that a line (unobstructed in antiquity) from Xian center (taking as reference point the Ming Bell Tower) to the highest point of the mountain cuts very neatly the huge mound of the Yiling Mausoleum of Emperor Ai Di (Liu Xin, 27B.C.-1B.C.) of the Han Dynasty, which was perhaps used as a guide to the eye.
Selecting a mountain for his tomb, emperor Taizong started a tradition which was to be followed by almost all the emperors of his dynasty. The first was the Emperor's son, Gaozong, who built the Qianling mausoleum, a magnificent interplay between nature and man-made architecture. Qianling is located at the foot of Liangshan Mountain, 1,049 meters above sea level. The main peak of the mountain is cone-shaped and resembles an artificial mound to which a magnificent *Spirit Path* – a paved road endowed with twin couples of ceremonial and apotropaic statues – approaches. As all others royal Tang tombs, the emperor's grave is unexcavated; we can, however, get a glimpse to how it may appear by visiting the tombs of Crown Prince Yide, Crown Prince Zhanghuai and Princess Yongtai which have been excavated in the necropolis associated to the main mausoleum (Eckfeld 2005, Wu 2010).
After Qianling, as much as 16 other Tang Emperors' Mausoleums were constructed in Guangzhong Plain, extending 150 kilometers from Qianling Mausoleum in the east to Tailing Mausoleum in the west. **FIGURE 1** Each emperor apparently asked to his experts to locate a suitable hill for the mausoleum, so that they distribute back and forth around the valley, taking advantage of suitably selected peaks. Not always were the lands in front of the peaks suitable as well for the construction of a Spirit Path of chosen direction, so that sometimes such paths depend on the topography. In other cases, their direction cannot be determined because the number of paired statues still present is not big enough. Using satellite images it is anyway possible to measure the axes of 13 mausoleums among 18, most of them with a reasonable accuracy. In a few cases the measures have also been controlled on site by the author using a precision magnetic compass (of course corrected for magnetic declination), while the data for Qiaoling and Tailing have also been cross-checked with existing high-precision surveys (Qiming and Koch 2002).
The azimuths, looking along the path in the direction of the mountain, are reported in Table 1. The following observations can be made:

1) The first mausoleum, as already mentioned, is actually a mound in Han style, and is oriented cardinally as are many Han mounds.

2) The orientation of the Spirit Path of the first mountain-based mausoleum, Taizong's, is unique in being towards the south (10° west of south). In other words, this is the unique Tang monument approached from the north. The reason is readily seen: the mountain raises in a very steep way on the plain to the south, while on the opposite side the approach is from an high plateau, with a sort of smooth valley in which the spirit path was built. Therefore, the origin of such an anomalous orientation is certainly topographical: it was obliged by the choice of the peak and the terrain around it. In a sense, it is a price the emperor had to pay in order to use the unique really prominent mountain visible from Xian. The proof is that this orientation was not used any more; in other words, all other spirit path of the Tang run from the plain towards their peaks to the north. The orientation of Taizong will not, therefore, be considered any more in what follows.

3) Two mausoleums have orientation at 348°. Inspection of satellite imagery shows that topographical considerations obliged this choice. The 7th (Janling) mausoleum is unique in that it actually has two, roughly parallel spirit paths (359° and 348°) which develop along two parallel natural reliefs that approach the mountain. Apparently then, the requirement of symmetry for the approaching path obliged the architect to build two of them. The case of the 6th mausoleum (Tailing) is similar, as the builders were obliged to respect the symmetry with respect to the surrounding hills.

4) The remaining measurable mausoleums exhibit orientations between 350° and 357°.

To understand these orientations, we recall that, in a recent paper, the Han mausoleums have been shown to be divided in two "families" (Magli 2018). Family 1 comprises monuments with a precise orientation to the cardinal points (with errors not exceeding ±1°); Family 2 comprises monuments roughly oriented to the cardinal points, but with errors in relation to the geographic north of several degrees. A behavior similar to that of Family 2 is exhibited by all the mausoleums of the (Northern) Song Dynasty (960-1127 AD), whose emperors revived the mound tradition.
The first family is clealry related to the pivotal role of the circumpolar stars and of the north celestial pole in the Chinese worldview, in which the ruler has the mandate of the heavens and in the heavens a counterpart of the celestial palace is located among the polar stars (Didier 2009a,b, Pankenier 2015). The second family might have been related to a compass orientation (Charvátová et al. 2011), but the evidence is very weak and we brought many new arguments which rather show that the orientation was to the maximal western elongation of the star Polaris. The maximal western elongation of a star is the distance in degrees between the star and the the pole measured when both are at the same height and the star is to the west (Chinese astronomers were accustomed to measuring this kind of stars' elongations, using a method called "the four excursions" based on a sighting tube fitted with a template and indentations; Needham 1959).
To understand why Polaris was chosen, we must first of all remember that, due to precession, the north celestial pole was in a dark region at those times. In fact, the pole had not been located sufficiently close to any bright "polar" star since the third millennium BC, when the pole star was Thuban, of the constellation Draco. Later the pole had been relatively close to the bright star Kochab, but in Han times it had already started to move towards Polaris. Actually, the decrease of its maximal western elongation can (roughly) be seen in agreement with a gradual shift in the orientations of the Han and the Song monuments. The symbolic reasons which were probably at the basis of the choice of Polaris have been discussed at length in Magli (2018): they reside in the symbolic role of the polar region of the sky in the Chinese imperial ideology of power (Needham 1959, Didier 2009c, Pankenier 2015).
Also in the case of the Tang mausoleums, a first alternative to be considered is a compass orientation according to compass Feng Shui, namely, the possibility that a compass was used to determine the most auspicious direction towards the mountain. To explore this issue, we calculated the correlation coefficient of azimuth vs. magnetic declination following the method described in general terms in the the previous section. The resulting value of R is 0.29. This very weak correlation value, taken together with the fact that all deviations are west of north, while the magnetic declination *changed sign* in that geographical area around 700 AD, shows that there is no correlation between the data and the secular variation of the magnetic field: compass Feng Shui was not used for the Tang mausoleums. By the way, the same actually holds for Form Feng Shui, since - if we let alone the obvious fact that the chosen mountains can be seen as "principal mountains" for the graves – no other typical features can be individuated.

So we are led to think that astronomical factors influenced the orientation of the Spirit Paths, exactly as occurred for the Han before and for the Song later on. During the Tang dynasty, the maximal western elongation of Polaris was ~350° in 600 AD reaching ~351.5° around 850 AD, in essential agreement with the data, which are all comprised between 350° and 357° and show a tendency to increase. Of course it is important to remember that these mausoleums are not man-made features but pre-existing hills and only the paths could be measured; this helps to explain the relative accuracy of the correspondence.

**4. The Ming tombs**

The Ming dynasty (1368-1644) was an exceptional period of stability of the Chinese empire. The first Ming capital was in Nanjing, where the tomb of the first emperor Hongwu – the Xiaoling Mausoleum – is located. In 1385 Hongwu also ordered the construction of an ancestor Mausoleum in Mingzuling town, Xuyi County (this monument was flooded in the 17 century and recovered in the last decades of the last century).

The successor Yongle moved the capital in Beijing. Consequently, for his tomb he had to choose a virgin place not far from Beijing, inaugurating a necropolis which was later used by all his successors. This necropolis is today known as *the Thirteen Tombs of the Ming Dynasty* and is located in Changping District, 42 kilometers to the northwest of the Forbidden City **FIGURE 2**. The complex is accessed trough a 7-kilometer Spirit path endowed with tens of statues and opened by a huge stone archway, completed in 1540. Single tombs are endowed with buildings and paths nestled to the main one. The plan of the burials is always similar: a huge rectangular enclosure terminating with a circular mound (an exception is the last tomb, which will be treated separately). Historical sources and modern authors all agree that Feng Shui was taken into account in the project of the royal tombs during the Ming dynasty (see e.g. Huadong 2013). A form Feng Shui inspiration can actually be seen already in the project of the first mausoleum, Xiao Ling. The tomb is located below the bell-shaped Mount Zhongshan, which therefore acts as the mountain to the north. Several hills extend southwards to form a "winding azure dragon" and westward as "tamed white tiger"; water flows to the south of the tomb, and the arch-shaped range of Mount Zijin may represent a "dragon vein."

The use of form feng Shui is apparent also in the choice of the thirteen Ming Tombs complex area. Indeed the first tomb is in the ideal midway of an arc-shaped valley open to the south and at the foot of Mount Tianshou to the north, whose 3 peaks tower directly on the axis of the tomb. The hills slope on both sides of this central mountain, and the setting is completed by the general approach to the site. Visitors are welcomed by the Red Gate, which is set in between two low hills to the east and the west respectively, thus playing the role of Dragon Hill and Tiger Hill. A branch of Wyniu river flows to the immediate south. An important characteristic of this site is that it actually allowed the construction of all the tombs in such a way that each one has a sort of "principal mountain" on its background. This Feng Shui requirement clealry governed orientation of each single tomb, so we do not expect a role for compass orientation. For the sake of completeness we performed the linear correlation test on the data reported in Table 2, which gives a lapidary R=0.05.

Clearly, we do not expect any role for astronomy either, but the special case of the last tomb, Siling, deserves attention.

The last Ming emperor Chongzhen committed suicide. Both historical sources and archaeological evidence show that the new rulers, the Qing, ordered his body to be buried with his concubine Tian, and it is reported that Tian's tomb was already existing and was on occasion declared as an imperial mausoleum. However, the topographical siting of this tomb is very strange. In fact, it is different not only from the royal tombs but also from all the other tombs of the Ming concubines (today much dilapidated and difficult to find) which are located in the southwestern part of the Necropolis. Their placement indeed, although less spectacular, is conceived in a similar manner as that of the emperors' tomb, so that they are orientated fronting the "veins" of the hills, although the area is certainly less favorable from Feng Shui viewpoint since the "principal mountains" do not lie to the north/north east but rather to the west. In any case, Siling is a glaring exception in being located in the full, flat plain and in front of a insignificant hill. In turn, its orientation is also a clear exception since it is quite precisely oriented to the geographical north, quite a good choice for an emperor tomb if it must be located in flat plain. So it is at least conceivable that, instead of searching attentively a favorable place for an emperor which was to be buried with honor but certainly was not the focus of attention of his successor, it was decided to build in flat plain and the builders opted for astronomical orientation.

## 5. The tombs of the Qing dynasty

5.1 Early Qing tombs

The Qing dynasty (1644–1912), the last imperial dynasty of China., originated in Manchu. Before the establishment of the capital in Beijing, the Qing thus built three imperial tombs north of the Great Wall: Yongling in Fushun, Fuling and Zhaoling in Shenyang (Table 3).
All these mausoleums are constructed in a quite peculiar style and endowed with elaborate sculptures; Yongling is the tomb of the Ancestors of the first Qing emperor, Nurhachi (who also built a further cemetery, Dongling in Liaoyang, for some of his relatives); Fuling is Nurhachi's tomb, and Zhaoling is the grave of the second emperor Hong Taiji. All these mausoleums are clealry oriented towards the rivers which flows directly to the south of them.

5.2 Qing Eastern Tombs

According to historians the last emperor of the Ming, Chongzhen, wanted to be buried in Zunhua, Hebei Province. To this aim he asked to individuate a favorable place there, and the geomancers choose the area called Fengtaling. Alas, as we have seen, the emperor did not use it. However, again according to the historians, the third emperor of the Qing Dynasty, Shunzhi (actually the first emperor to rule over the whole of China) was himself an expert of Feng Shui. After moving the capital to Beijing, he started searching for a place for his tomb. When he visited Fengtaling, he confirmed the auspiciousness of the area and asked for his tomb to be placed there. In this way he inaugurated what is today called the necropolis of the eastern Qing tombs, where 4 emperors, 4 empresses, and more than 150 members of the royal family are buried (**FIGURE 3**).
In spite of the above mentioned, repeated claims of auspiciousness present in the written sources, the general shape of the Fengtaling area - although suggestive - is quite different from that of the Ming tombs where, as we have seen, all the elements of form Feng Shui can be easily individuated. Looking at Fengtaling it is immediately seen that the northern mountain – although shaped in a "dragon" profile - does not encircle the valley but it is rather a straight front-line of hills, the Yanshan Mountains, with the front having an approximate azimuth of 58°. To the far south, a small "protective" hill, Mount Jinxing, can be seen. Overall, the place appears to be especially suited only for the first tomb, Xiaoling, which is placed right at the foot of the main peak (Mount Changrui) and on the central axis of the site. The other tombs are spread on both sides of the main one but lack a specific "mountain", and so it is understandable that doubts could arise on the opportunity of adding tombs to the same site, a factor which may have influenced the foundation of the Western necropolis (see below).

The orientations of the tombs (Table 4) are readily seen to be all roughly orthogonal to the hill's front line. In any case, we have performed the correlation coefficient test against the behavior of the magnetic declination, which returns a negative value (R=-0.71) showing – for the reasons discussed in section 2 - complete absence of correlation.

5.3 Qing Western Tombs

After Shunzhi, also his son Kangxi was buried in the eastern tombs cemetery. However Kangxi's son, Yongzheng (1678–1735), decided to change place and moved to Yixian, founding the necropolis today known as the Western Qing tombs (**FIGURE 4**).

It is not clear why this emperor made the choice of changing burial site; an explanation evokes the alleged fact that he murdered his father, so that he feared being buried next to him. This is, however, difficult to believe, and another explanation might be that a tomb was begun for him but then the soil revealed inauspicious and humid, and he interpreted this as a sign that a radical change was needed. A construction place that was clealry meant to become a royal tomb and was left unfinished, with only the shape of the enclosure-mound structure delineated, is visible in the easternmost area of the eastern cemetery. In existing literature however, this is usually attributed to the later emperor Daugang, who is known for certain to have abandoned a construction site in favor of the Western Tombs. No material evidence is, however, cited for this attribution, and the fact that the abandoned project is roughly parallel and immediately to the east to the last tomb of the Kangxi group, casts doubts about this attribution.

Whatever the reason, a new necropolis was thus founded by Yongzheng. This necropolis, today known as the Western Qing Tombs, is located in Yi County, Hebei. After being inaugurated, the necropolis was used by four emperors: Jiaqing (1760–1820, the 5th emperor) Daoguang (1782–1850, the 6th emperor) and Guangxu (1871–1908, the 9th emperor).

The place chosen is certainly beautiful and and is "protected" from the north by a far mountain range (the Yongnilg Mountains). An impressively beautiful forest of ancient pines surrounds the place still today, but it is difficult to say which was the extension of pine forests hundreds of years ago (and so, if the site might have been individuated essentially for this reason). Besides vegetation, it is very difficult to identify other typical Feng Shui auspicious features. Actually, the tombs are scattered at the basis of the low hills which border the valley essentially in a continuous way for several tens of kilometers, and no "womb like" morphology can be individuated anywhere. As far as orientation is concerned, each tomb more or less points along the direction of maximal steep of the (usually low) hill to its immediate north (Table 5). We anyhow performed the test for magnetic correlation which returns a negative value R=-0.12.

At least in the present author's view, the reasons for the choice of the site for the western necropolis remain obscure and cannot be ascribed to Feng Shui, or at least to feng Shui only. In this respect, an observation can however be made, although with a warning to the reader that the fact may be purely due to a chance. The distance as the crow flies between the center of the Forbidden city in Beijing and the first mausoleum built in the eastern necropolis is of 111.8 km, a value very close to the distance calculated from the same point and the first tomb built in the western necropolis, which is 108,4 km (the difference is less than 3%). Of course the two sites are not inter-visible, but precise geodetic measures were certainly within the skills of the Chinese geographers of the period. This raises the possibility that, once decided to found a new necropolis, the first of the western Qing tombs was built in the spot were the circumference centered in the forbidden city intersects the border between the valley and the hills. Interestingly, taking into account also the position of the Ming tombs this creates a (roughly) symmetrical configuration illustrated in **FIGURE 5** (the two angles are of 91° to the east and of 106° to the west).

# 6. Discussion and conclusions

In the present paper, we analyzed the royal necropolis of the Tang, Ming and Qing dynasties in an effort to clarify cognitive aspects related to orientation and placement in the landscape. In particular, we wanted to investigate the possible role of astronomy and of traditional Feng Shui doctrine, both in the "form" and in the "compass" version.

The first conclusion we can definitively draw is that compass Feng Shui had *no role* in the planning of these tombs. This is shown quantitatively by the lack of positive correlation between orientation data and variation of the magnetic declination. Together with the results of Magli (2018) this completes in the negative the analysis about the use of compass orientation in the main necropolises of the Chinese emperors.

A second result of this paper is that the orientation of the Tang tombs almost certainly followed the tradition inaugurated by the western Han mausoleums and later followed by the Song ones, namely astronomical orientation towards the western elongation of Polaris.

This tradition was lost with the Ming dynasty: during the Ming rule the Form Feng Shui doctrine became fundamental for the architectural choices. This latter fact was, of course, already well known; however, it is confirmed here by excluding quantitatively astronomical and magnetic orientation. Finally, regarding the Qing dynasty a distinction appears necessary between the Eastern necropolis and the Western one. It is in fact conceivable that form Feng Shui inspired the location of the first, while the same cannot be said about the second. We thus tentatively proposed a "cartographic" placement for the latter.

**Acknowledgments**

The author thanks the GFZ group on Geomagnetism, and in particular Claudia Stolle, Monika Korte and Sania Panovska, for making their models freely available and for their advice in applying them. Eugenio Realini and Giulio Tagliaferro are gratefully acknowledged for their help in producing the data used in this paper. Finally, the author thanks Guido Heinz and Susanne Greiff at Römisch-Germanisches Zentralmuseum Mainz for making available their surveys of Mausoleums in Pucheng County.

| Emperor | Tomb's Name | Accession | Az. | Magnetic dec. | Notes |
|---|---|---|---|---|---|
| 1 Gaozu | Xianling | 618 | 359 | 1.32 | Mound |
| 2 Taizong | Zhaoling | 626 | 190 | 1.32 | Spirit path from the north |
| 3 Gao Zong | Qianling , | 649 | 354 | 1.16 | |
| 4 Zhongzong | Dingling | 705 | | -0.12 | Non meas. |
| 5 Ruizong | Qiaoling | 684 | 356 | 0.49 | |
| 6 Xuanzong | Tailing | 712 | 348 | -0.29 | Topographical |
| 7 Suzong | Jianling | 756 | 359 and 348 | -2.18 | Topographical |
| 8 Daizong | Yuanling | 762 | | -2.41 | Non meas. |
| 9 Dezong | Chongling | 779 | | -3.34 | Non meas. |
| 10 Shunzong | Fengling | 805 | 350 | -4.45 | |
| 11 Xianzong | Jingling | 805 | 352 | -4.45 | |
| 12 Muzong | Guangling | 820 | 353 | -5.05 | |
| 13 Jingzong | Zhuangling | 824 | 356 | -5.23 | |
| 14 Wenzong | Zhangling | 827 | 355 | -5.23 | |
| 15 Wuzong | Ruiling | 840 | 355 | -5.72 | |
| 16 Xuanzong | Zhenling | 846 | | -5.86 | Non meas. |
| 17 Yizong | Jianling | 859 | 358 | -6.22 | |
| 18 Xizong | Jingling . | 873 | | -6.48 | Non meas. |

**Table 1**
**Mausoleums of the Tang dynasty**

| Emperor | Name of Tomb | Accession | Az. | Magnetic decl. | NOTES |
|---|---|---|---|---|---|
| Ming Ancestors | Ming Zuling | 1368 | 2 | 0.32 | Huaian; built by Hongwu |
| Hongwu | Xiaoling | 1368 | 0° | 0.26 | Nanjing |
| Yongle | Chang Ling | 1402 | 6 | -0.36 | |
| Hongxi | Xian Ling | 1424 | 16 | -0.83 | |
| Xuande | Jing Ling | 1425 | 51 | -0.83 | |
| Zhengtong | Yu Ling | 1435 | 17 | -1.00 | |
| Chenghua | Mao Ling | 1464 | 11 | -1.42 | |
| 6Hongzhi | Tai Ling | 1487 | 346 | -1.58 | |
| Zhengde | Kang Ling | 1505 | 292 | -1.62 | |
| Jiajing | Yong Ling | 1521 | 50 | -1.58 | |
| Longqing | Zhao Ling | 1567 | 320 | -1.25 | |
| Wanli | Dingling | 1572 | 299 | -1.22 | |
| Taichang | Qing Ling | 1620 | 14 | -1.26 | |
| Tianqi | De Ling | 1620 | 84 | -1.26 | |
| Chongzhen | Si Ling | 1627 | 358 | -1.35 | |
| Gongruixian (Jiajing's father) | Xianling | 1521 | 31 | -1.62 | Zhongxiang |

**Table 2**
**Mausoleums of the Ming dynasty**

| Emperor | Name of tomb | Accession date | Az. | Magnetic Decl. | Notes |
|---|---|---|---|---|---|
| **Parents of Nurhaci** | Dongjing | 1624 | 324 and 64 | -0.80 | Liaoyang |
| **Ancestor's of Qing dyansty** | Yongling | 1598 | 318 | -0.70 | Fushun |
| **Nurhaci** | Fuling | 1616 | 344 | -0.75 | Shenyang |
| **Hong Taji** | Zhaoling | 1626 | 357 | -0.80 | Shenyang |

**Table 3
Early Qing tombs**

| label | emperor | Name of tomb | Accessionj date | azimuth | Magnetic declination | notes |
|---|---|---|---|---|---|---|
| **1** | **Shunzhi** | **Xiaoling,** | **1644** | **334** | **-1.05** | |
| **1A** | | | | 332 | | wife |
| **1 B** | | | | 345 | | mother |
| **2** | **Kangxi** | **Jingling,** | **1662** | **354** | **-1.19** | |
| **2 A** | | | | 349 | | |
| **2 B** | | | | 354 | | |
| | Yongzheng | | | | | West tombs |
| **3** | **Qianlong** | **Yuling** | **1736** | **329** | **-1.22** | |
| **3 A** | | | | 345 | | |
| Jiaqing | Jiaqing | | | | | West tombs |
| Daoguang | Daoguang | | | | | West tombs |
| **4** | **Xianfeng** | **Dingling** | **1850** | **348** | **-2.41** | |
| **4 A** | | | | 358 | | |
| **4 B** | | **Ding Dongling** | | 336 | | **Twin tomb for Empress Dowager Cixi and Empress Dowager Cian** |
| **5** | **Tongzhi** | **Huiling** | **1861** | **9** | **-2.64** | |
| **5 A** | | | | 5 | | |
| Guangxu | | | | | | West tombs |
| **6** | | | | 352 | | **Abandoned construction site usually attributed to the Daoguang Emperor.** |
| **6 A** | | | | 4 | | **Two sons and two daughters of Daoguang** |

**Table 4**
**Mausoleums of the Eastern Qing tombs**
**(for clarity, all the Qing emperors are listed, also those buried in the western cemetery)**

| Label | emperor | Name of tomb | Acc date | Azimuth | Magnetic declination | Note |
|---|---|---|---|---|---|---|
| 1 | Yongzheng | Tailing | 1723 | 347 | -1.34 | |
| 1a | | | | 14 | | 21 concubines |
| 1 b | | | | 9 | | |
| 2 | Jiaqing | Changling | 1796 | 354 | -1.31 | |
| 2a | | | | 352 | | 17 concubines |
| 2 B | | | | 17 | | |
| 2 c | | | | 13 | | 8th son |
| 3 | Daoguang | Muling | 1821 | 302 | -1.33 | |
| 3 a | | | | 7 | | wife |
| 4 | Guangxu | Chongling | 1875 | 335 | -2.15 | |
| 4 a | | | | 20 | | two concubines |
| 4 b | | | | 11 | | |
| 4 c | | | | 12 | | |
| 4 d | | | | 352 | | |

**Table 5**
**Mausoleums of the western Qing tombs**

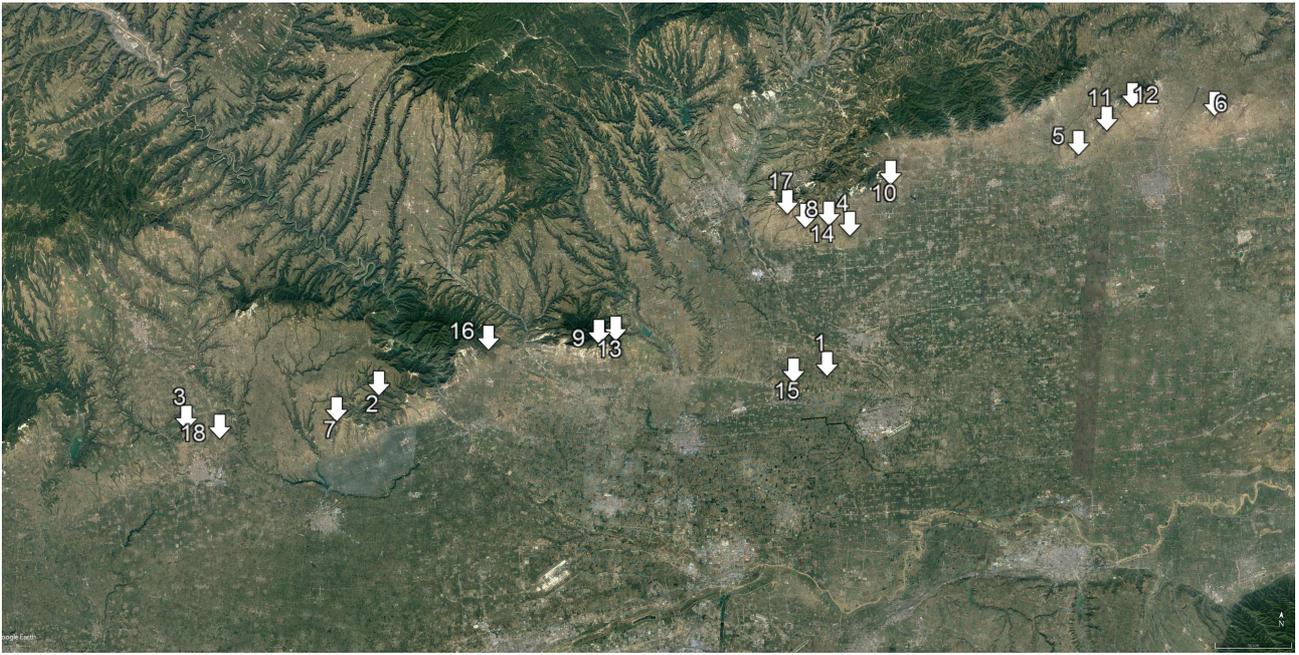

**Fig. 1**
The Tang mausoleums. 1 Gaozu Xianling 2 Taizong Zhaoling 3 Gao Zong Qianling, 4 Zhongzong Dingling 5 Ruizong Qiaoling 6 Xuanzong Tailing 7 Suzong Jianling 8 Daizong Yuanling 9 Dezong Chongling 10 Shunzong Fengling 11 Xianzong Jingling 12 Muzong Guangling 13 Jingzong Zhuangling 14 Wenzong Zhangling 15 Wuzong Ruiling 16 Xuanzong Zhenling 17 Yizong Jianling 18 Xizong Jingling. (Image courtesy Google Earth, editing by the author)

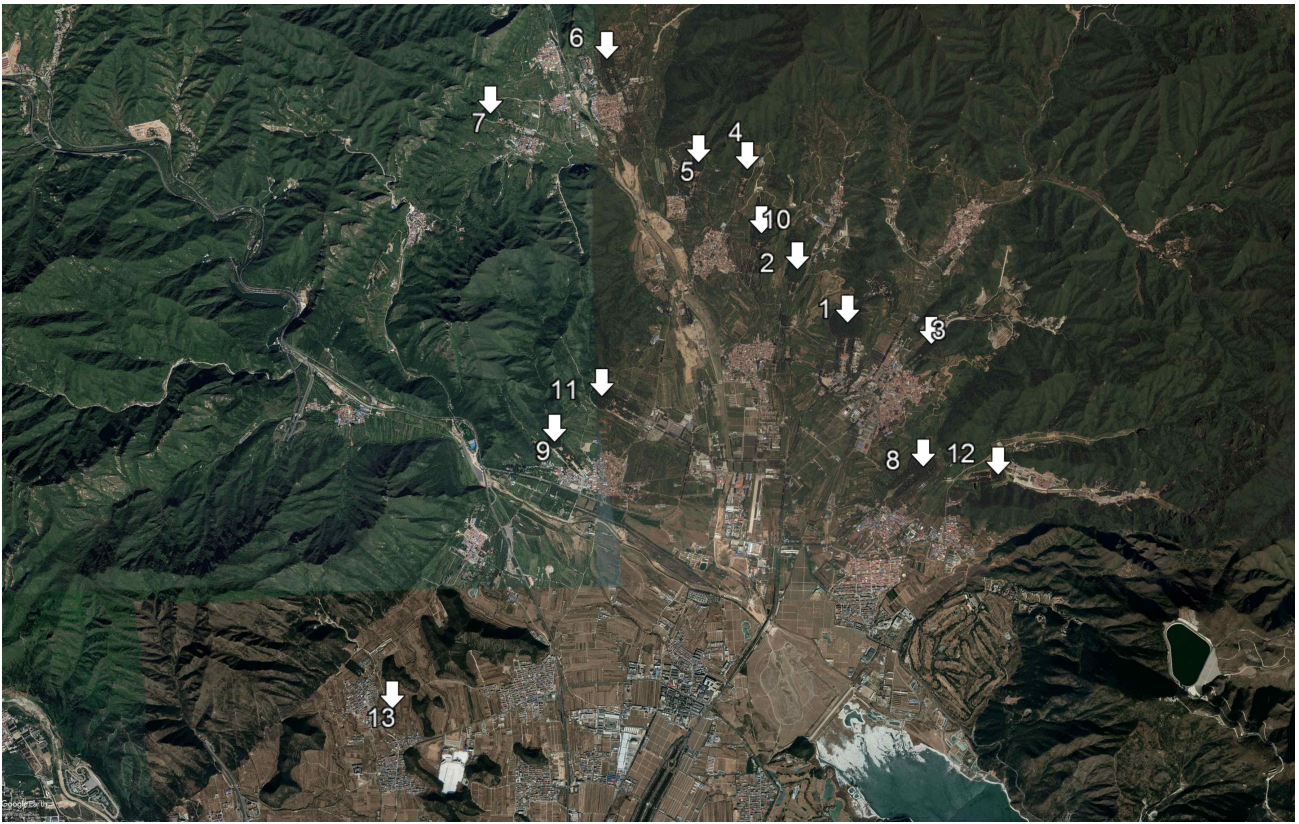

**Fig. 2**
**The Thirteen Ming tombs. 1 Yongle Chang Ling 2 Hongxi Xianling 3 Xuande Jingling 4 Zhengtong Yuling 5 Chenghua Maoling 6 Hongzhi Tailing 7 Zhengde Kangling 8 Jiajing Yongling 9 Longqing Zhaoling 10 Wanli Dingling 11 Taichang Qingling 12 Tianqi Deling 13 Chongzhen Siling. (Image courtesy Google Earth, editing by the author)**

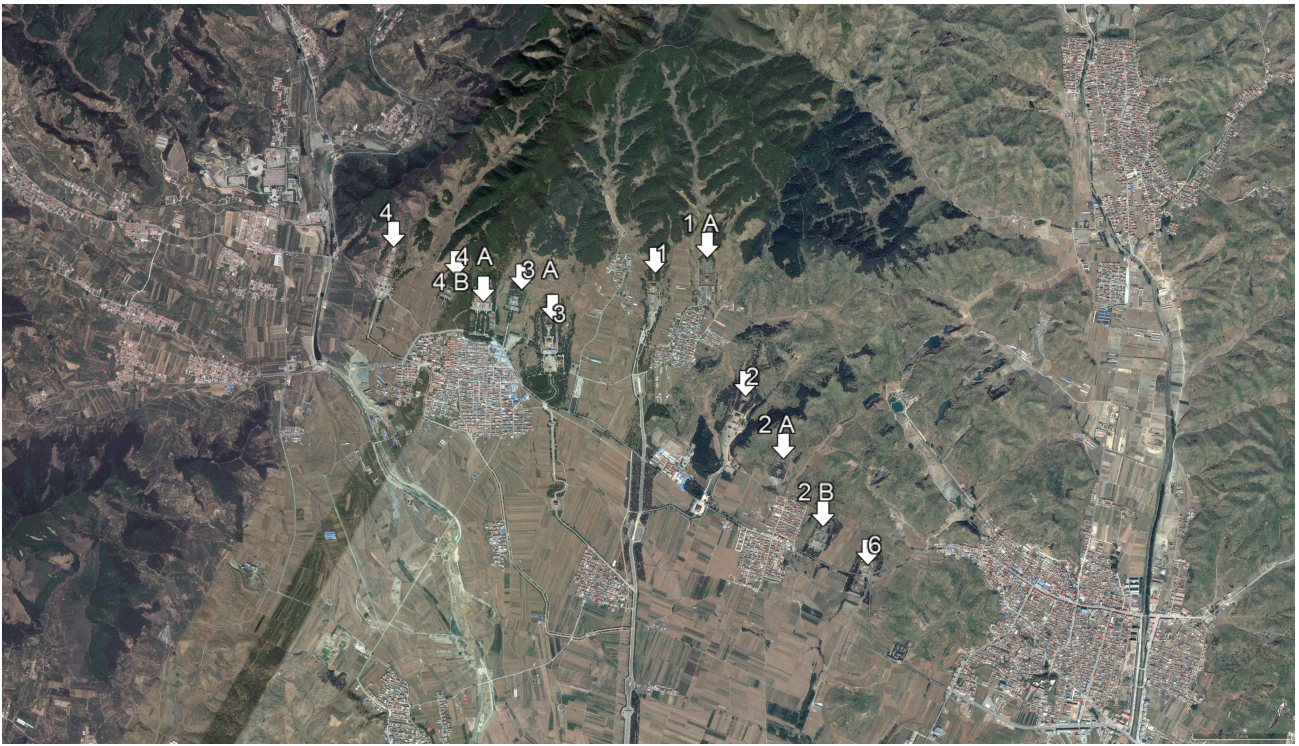

**Fig. 3**
**The imperial tombs and their satellites in the Eastern Qing tombs area. 1 Shunzhi Xiaoling 2 Kangxi Jingling 3 Qianlong Yuling 4 Xianfeng Dingling 5 Tongzhi Huiling 6 Unfinished (Daoguang?). Notice that north is slightly skewed to the right of top here to enhance the topographical features.
(Image courtesy Google Earth, editing by the author)**

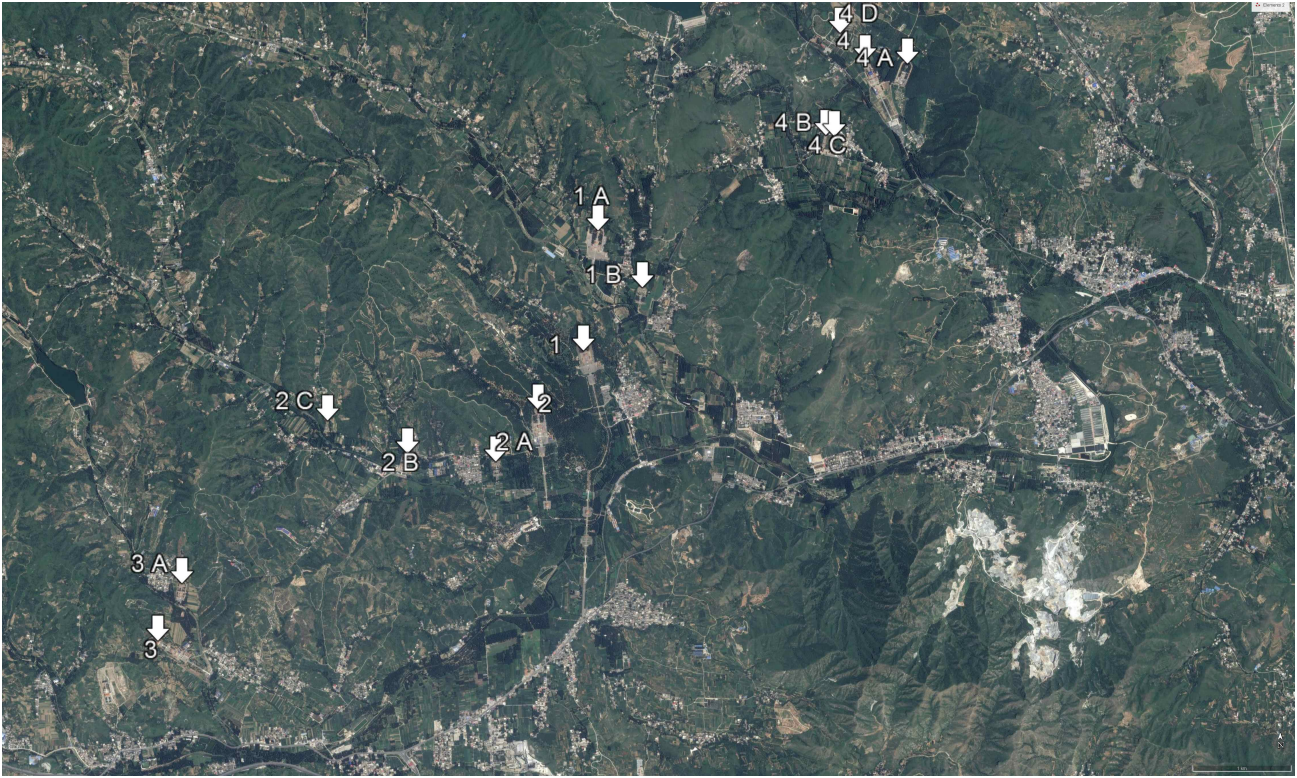

**Fig. 4**
**The imperial tombs and their satellites in the Western Qing tombs area. 1 Yongzheng Tailing 2 Jiaqing Changling 3 Daoguang Muling 4 Guangxu Chongling. (Image courtesy Google Earth, editing by the author)**

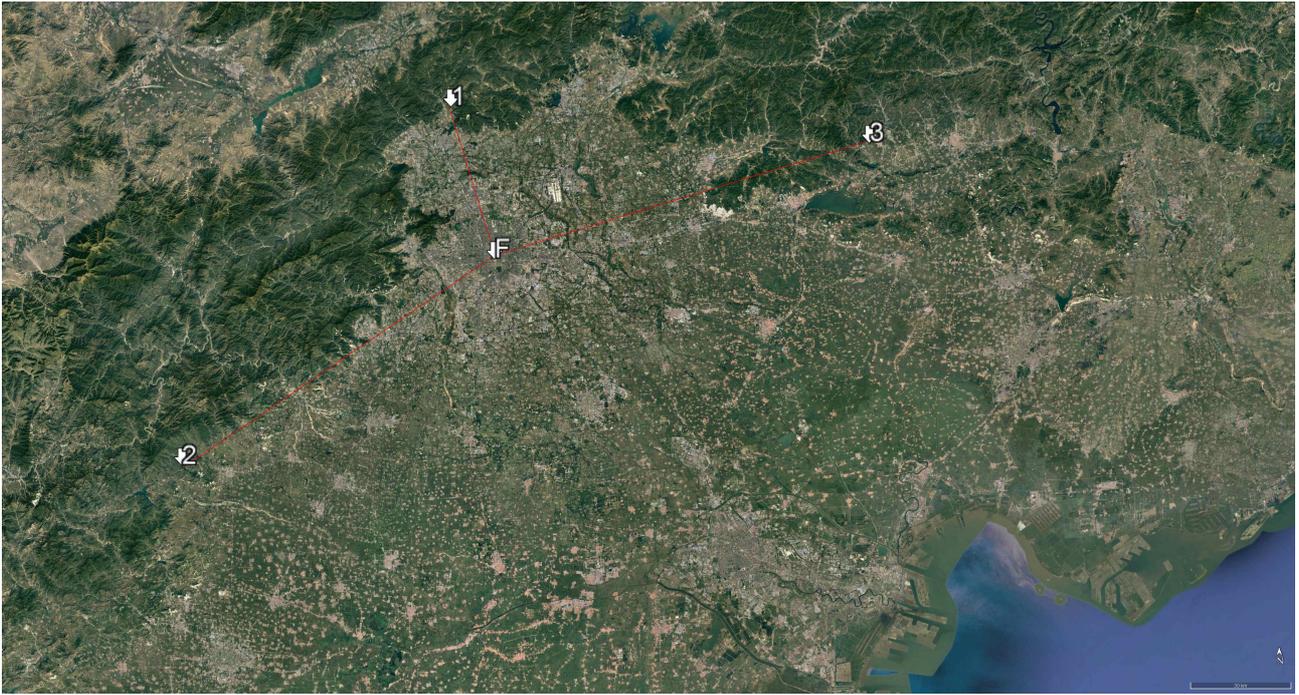

**Fig. 5**
**The positions of the Ming tombs 1 and of the Western 2 and Eastern 3 Qing tombs with respect to the Forbidden City (F) in Beijing. (Image courtesy Google Earth, editing by the author)**